\documentclass[aps,twocolumn,superscriptaddress,amsmath,amssymb]{revtex4-1}
 \usepackage{soul}
\usepackage{graphics}
\usepackage{epsfig}
\usepackage{dcolumn}
\usepackage{color}
\usepackage{bm}
\usepackage{subfigure}
\usepackage{hyperref}
\usepackage[mathlines]{lineno}
\usepackage{amsmath}
\usepackage{amssymb}
\usepackage{comment}

\begin{document}

\title{Non-local sidewall response and deviation from exact quantization of the topological magnetoelectric effect in axion-insulator thin film}
\author{N.~Pournaghavi}
\affiliation{Department of Physics and Electrical Engineering, Linn{\ae}us University, 391 82 Kalmar, Sweden}
\author{A.~Pertsova}
\affiliation{Nordita, KTH Royal Institute of Technology and Stockholm University, Roslagstullsbacken 23, SE-106 91 Stockholm, Sweden}
\author{A.~H.~MacDonald}
\affiliation{Department of Physics, University of Texas at Austin, TX 78712, USA}
\author{C.~M.~Canali}
\affiliation{Department of Physics and Electrical Engineering, Linn{\ae}us University, 391 82 Kalmar, Sweden}

\begin{abstract}
Topological insulator (TI) thin films with surface magnetism are expected to exhibit a quantized anomalous Hall effect (QAHE) when the magnetizations on the top and bottom surfaces are parallel, and a quantized topological magnetoelectric (QTME) response when the magnetizations have opposing orientations (axion insulator phase) and the films are sufficiently thick. We present a unified picture of both effects that associates deviations from exact quantization of the QTME caused by finite thickness with non-locality in the side-wall current response function. Using realistic tight-binding model calculations, we show that in $Bi_2Se_3$ TI thin films deviations from quantization in the axion insulator-phase are reduced in size when the exchange coupling of tight-binding model basis states to the local magnetization near the surface is strengthened. Stronger exchange coupling also reduces the effect of potential disorder, which is unimportant for the QAHE but detrimental for the QTME, which requires that the Fermi energy lie inside the gap at all positions.
\end{abstract}


\maketitle

\noindent
\textit{Introduction}--- In magnetoelectric materials, 
an applied electric field
$\bf E$ induces magnetization $\bf M$ and an applied magnetic field $\bf B$
induces electrical polarization $\bf P$\cite{Fiebig2005}. Magnetoelectric 
response is described by the linear magnetoelectric polarizability 
tensor $\alpha$, whose diagonal components,
\begin{equation}
\alpha_{ii} = \frac{\partial M_i}{\partial E_i}\Bigg\vert_{{\bf B}= 0} = 
\  \frac{\partial P_i}{\partial B_i}\Bigg\vert_{{\bf E}=0} = 
\frac{\theta}{2\pi}\frac{e^2}{h}\;,
\label{me_tensor}
\end{equation}
are pseudoscalars.  ($\theta$ in Eq.~\ref{me_tensor} is dimensionless.)
Since $\bf B$ and $\bf M$ are odd under time-reversal  
and $\bf E$ and $\bf P$ are odd under
space-inversion, 
magnetoelectric response normally occurs in  
insulators that break both time-reversal and inversion symmetry (TRS and IS), 
and is typically characterized by a small value of $\theta$.
We are interested here in the orbital magnetoelectric response \cite{Qi_PRB2008, essin_PRL2009, malaschevich2010} 
of three-dimensional (3D) topological insulators (TIs) \cite{Hasan2010,XLQi},
which is special in the sense that it is non-zero even when TRS
is not broken in the sample bulk\cite{armitage2018}.  
The magneto-electric response is instead related to a non-trivial 
topological invariant of the bulk bands \cite{Qi_PRB2008, essin_PRL2009, XLQi} 
and is quantized at $\theta=\pi$ ($\alpha_{ii} = e^2/2h $).

\begin{figure}[ht!]
\centering
\includegraphics[width=0.98\linewidth,clip=true]{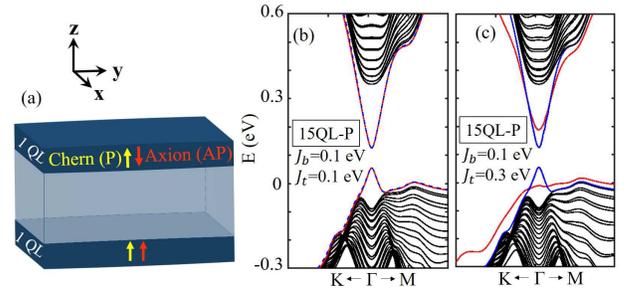}
\caption{(a) Schematic of a TI thin film with surface 
magnetizations at top and bottom in 
either parallel (P: Chern-insulator phase) or antiparallel (AP: axion-insulator phase) configurations.
(b)-(c) Bandstructure of a 15QL Bi$_2$Se$_3$ TI thin film in the P 
configuration for two choices of the top and bottom surface exchange fields $J_b$ and $J_t$. 
The gapped Dirac surface states are indicated by red and blue lines.
In the symmetric case (b) these states are nearly doubly degenerate, and 
approach exact degeneracy in the thick film limit.
} 
\label{fig1}
\end{figure}

The Quantized Topological Magnetoelectric Effect (QTME) is realized only when the TI surface magnetization
adopts an axion insulator configuration \cite{Qi_PRB2008,essin_PRL2009}, one in which  
all facets and hinges (where facets meet) of the bulk TI crystal surface are insulating\footnote{If only the top and bottom surfaces were magnetized, the sidewalls of a finite heterostructure would not be insulating since they would host conducting surface states.  These states can however be gapped by i) designing geometries in which an orthogonal the exchange field is also present on the sidewall or ii) exploiting energy level quantization by tapering off the thin film thickness near the sidewalls.}.  For thin films this requires that the top and bottom surface magnetizations have opposite orientations \cite{wang_PRB2015, morimoto_PRB2015}, 
as shown in Fig.~\ref{fig1}. When the magnetizations have the same orientation, the film displays \cite{Yu2010} the Quantum Anomalous Hall Effect (QAHE) and the side walls 
(hinges) are not gapped.  Both the QAHE and the QTME can be understood qualitatively \cite{wang_PRB2015} by considering the limit of weakly-gapped Dirac cone surface 
states, since these give rise to half-quantized intrinsic anomalous Hall conductances $\sigma_{\rm H} = \pm e^2/2h$
with a sign determined by the magnetization orientation\cite{Fu_Kane2007,Qi_PRB2008,XLQi}.  
The QAHE has been observed in uniformly-doped magnetic TI thin films\cite{chang2013},
in modulation-doped TI films \cite{mogi_APL2015,mogi_NatMat2017, mogi_ScAdv2017,xiao2018, Allen2019} that
have surface magnetism only, and recently also in TI films with proximity-induced 2D magnetism\cite{mogi_APL2019}. 
On the other hand, the QTME not yet been directly \cite{QuantumFaraday} confirmed experimentally, even though successful realization
of the axion insulator configuration is strongly suggested in some experiments \cite{mogi_NatMat2017, mogi_ScAdv2017,xiao2018,he_PRL2018, he_PRL2018, Allen2019}
by the absence of a Hall effect in states with oppositely oriented top and bottom surface magnetizations. Novel intrinsic antiferromagnetic TIs, such as the van der Waals layered MnBi$_2$Te$_4$\cite{Otrokov2017,Li2019,Otrokov2019} and Mn$_4$Bi$_2$Te$_7$ families  \cite{Hirahara2020} and their heterostructures with nonmagnetic TIs\cite{Hirahara2017, Hirahara2020} have also been found to display the QAHE. These systems do not suffer from the intrinsic disorder of doped TIs and typically possess relatively larger magnetic gaps at their Dirac points.

In this Letter we employ a unified description of the QAHE and the QTME in TI thin films,
by expressing the magnetization in terms of sidewall currents that respond non-locally to electric potentials that vary slowly across the film. 
In this picture, perfect quantization of the QTME requires sidewall response that is localized 
near the top and bottom surfaces, whereas the QAHE requires only bulk state localization. 
We characterize the non-locality of the sidewall response by calculating finite-size corrections to the QTME theoretically using a realistic tight-binding model, 
demonstrating that they are smaller for stronger exchange coupling $J({\bf r})$
between the surface magnetization and tight-binding model basis states localized near the surface.  
By increasing surface state gaps, stronger exchange coupling not only reduces finite-size corrections but also reduces the effect of disorder, which can be tolerated in QAHE measurements but is deleterious for the QTME.

\textit{QAHE, QTME, and sidewall response}---
We consider the linear response of the $\hat{z}$-direction orbital 
magnetization of a thin film with a quasi-2D bulk gap to an electric potential that varies slowly across the film.  Since the bulk is time-reversal invariant and insulating,
the magnetization response must originate from changes in currents that circulate around the film side walls: 
$ M_z = (1/d) \sum_l I_{\rm sw}(l) $.
Here $d$ is the thickness of the film, 
and we have anticipated our use of a tight-binding model by specifying the vertical 
position along the side wall using a discrete layer index $l$.
By allowing the $l$-dependent bulk electric potentials to turn on slowly 
upon entering the sample bulk \cite{Xiao2010} so that lateral electric fields 
are present only near the sidewall, we can relate the current response to bulk Hall conductivity:
\begin{equation} 
I_{\rm sw}(l) = \frac{1}{e} \sum_{l'} \sigma_H(l,l') (V(l')-\mu)
\label{eq:swnl}
\end{equation} 
where $V(l)$ is a layer-dependent electric potential, and $\sigma_H(l,l')$ 
is the thin-film Hall conductivity generalized \cite{caveat1} to allow for non-locality in the $\hat{z}$-direction~\cite{MacDonald2021}: 
\begin{widetext} 
\begin{equation}
\sigma_H(l,l') = \frac{2e^2}{\hbar} \sum_{n'\ne n} f_n \int \frac{d^{2}\bm{k}}{(2\pi)^2} \; 
\frac{{\rm Im}[\langle u_{n{\bf k}}| P_l v_x ({\bf k}) |u_{n'{\bf k}}\rangle
\langle u_{n'{\bf k}}| P_{l'} v_y({\bf k}) |u_{n{\bf k}}\rangle ]}
{(E_{n{\bf k}} -  E_{n'{\bf k}}) ^2}.
\label{eq:nonlocalHall}
\end{equation}
\end{widetext}
In Eq.~\ref{eq:swnl} we have introduced a chemical potential to allow for a unified discussion of the 
QAHE and QTME.  In Eq.~\ref{eq:nonlocalHall} $f_n$ is a band occupation number, $P_l$ is a layer projection operator, $|u_{n{\bf k}}\rangle$ is a band state of the 2D Bloch Hamiltonian $H({\bf k})$, $E_{n{\bf k}}$ is the corresponding band energy and $v_i({\bf k}) = \partial H({\bf k})/ \partial k_i, \ i= x, y$ is the velocity operator.
When summed over $l$ and $l'$, $\sigma_{H}(l,l')$ yields $e^2/h$ times the total Chern number $C$ of all 
occupied 2D-bands, and is quantized.

The QAHE measures \cite{macdonald1995mesoscopic} the response of the total sidewall current to a uniform 
chemical potential shift, and its quantization is therefore explained simply by the 
quasi-2D band Chern numbers.  The QTME is a zero-temperature property of a state that is 
fully insulating, and is a response not to chemical potential but to electric potential.
Its quantization can nevertheless be understood in terms of Chern quantization by the following argument. Define $ \sigma_{t(b)}= \sum_{l,l'\in\ t(b)} \sigma_{H}(l,l')$, where each layer is classified
by proximity as belonging to the top or bottom layer subset.  For thick films the side-wall
response must be localized where time-reversal symmetry is broken, {\it i.e.} near the top or bottom surface.
It follows that for any configuration of the surface magnetism 
$\sigma_H=\sigma_t+\sigma_b$.  Since $\sigma_H$ is quantized, its value must be independent of 
small variations in local properties, including variations in the strength of the exchange coupling, 
which occur only at one surface and can change only $\sigma_{t}$ or $\sigma_{b}$.  
It follows that $\sigma_t$ and $\sigma_b$ must be separately universal.  Since both
must change sign when the magnetization is reversed at their surface, $\pm \sigma_t \pm \sigma_b$ must be quantized.
In the special case of an axion insulator ($\sigma_H=0$) it follows that $\sigma_b=-\sigma_t$,
$2 \sigma_t = n e^2/h$, 
and that for $n=1$, $\delta M \equiv M(E_z) - M(0)\approx (1/e d) \sigma_{t} (V(l_t)- V(l_b)) =
(e^2/2 h) E_z$
when an electric field $E_z$ is applied across the sample in the insulating state. 
$V(l_{t (b)})$ is the electric potential at the top (bottom) layer with layer index
$l_{t(b)}$. (Note that for asymmetric exchange fields at the two surfaces, the magnetization is non-zero even when  $E_z= 0$.)
Because the side-wall response has a finite localization length, the magnetization response 
has a finite size correction that varies inversely with the number of layers $N$ in the film and
is characterized by the dimensionless number
$m_{\rm corr}(d)  \equiv |\big (\delta M(d = \infty) - \delta M(d)\big)/ \delta M(d = \infty)|
= (h/4e^2) \sum_{(l') \in t} \sigma_H (l_t-l')/N$.

\textit{Finite Thickness Corrections in Bi$_2$Se$_3$}---
Finite-size corrections in thin films distinguish the QTME from the QAHE. To estimate their size in realistic systems
we have added a uniform electric field $E_z$ applied across
finite thickness quasi-2D TI films to
a realistic tight-binding (TB) model with surface magnetism, 
and explicitly evaluated the magnetization
carried by the distorted bands using \cite{Thonhauser2005,Xiao2005} 
\begin{widetext}
\begin{equation}
{M_z}(E_z)  = - \frac{e}{\hbar}\;  \sum_{n, n'} f_n
\int_{\rm BZ} \frac{d^2 k}{(2\pi)^2} \frac{( E_{n{\bf k}} +  E_{n'{\bf k}} -2 \mu)}
{( E_{n{\bf k}} -  E_{n'{\bf k}}) ^2}
 {\rm Im}\Big[ \langle u_{n{\bf k}} |v_y({\bf k}) |u_{n'{\bf k}}\rangle \langle u_{n'{\bf k}} |{v}_x({\bf k})| u_{n{\bf k}}\rangle
\Big]\;.
\label{final2}
\end{equation}
\end{widetext} 
Note that Eq.~\ref{final2} and Eq.~\ref{eq:swnl} agree in the case of constant $V(l)$ since a constant electric potential shifts band energies without changing wavefunctions. Eq.~\ref{final2} gives the 2D bulk magnetization for an {\it infinite} crossectional area 
TI slab with broken TRS at the surfaces.  The physical origin of the response of this magnetization to an electric field is the changes in the side-wall currents discussed above. 
This example of bulk-edge correspondence is closely analogous to that of the QAHE.\cite{macdonald1995mesoscopic, streda1982}

We focus on Bi$_2$Se$_3$ thin films~\cite{Hsieh2009}, whose
electronic structure can be described by a \textit{sp}$^3$ TB model
with parameters obtained by fitting to \textit{ab initio} electronic structure
calculations~\cite{Kobayashi2011,Pertsova2014}.
We apply this TB model to a thin-films with finite numbers of van-der-Waals-coupled quintuple layers (QLs),
and model broken time-reversal at the top and bottom surfaces by adding exchange fields
of strength $J_{\rm t}$ and $J_{\rm b}$, 
oriented orthogonal to the ($111$) surface that couple to electron spin.
We will consider two types of exchange-fields: 
(i) a homogeneous field applied to the entire first surface QL, modelling magnetic modulation doping\cite{mogi_APL2015, mogi_ScAdv2017,mogi_NatMat2017, xiao2018, Allen2019}, and 
(ii) a homogeneous field applied only to the very top
and bottom atomic monolayers (MLs), modelling the exponentially 
evanescent proximity effect of an adjacent magnetic layer\cite{he_PRL2018, mogi_APL2019, mogi_PRL2019}.

In Figs.~\ref{fig1}(b), (c) we plot the bandstructures of 15QL TI films for two
strengths of exchange fields of type (i), both in the parallel (P) QAHE configuration, corresponding to the Chern insulator phase. 
For symmetric exchange fields  
$J_{\rm t(b)} = 0.1$ eV (smaller than the bulk gap $\approx 0.3$ eV), 
the in-gap states on the two surfaces are essentially degenerate Dirac cones with exchange gaps
$\Delta \approx J$ at the DP.  In the following, we will refer to the surface states below the exchange 
gap as valence-band states and to those above as conduction-band states.
An analysis of the wavefunctions\cite{SM} shows that, around the 
$\Gamma$ point, these states are localized either at the top or bottom surface,
decaying exponentially within the first two QLs, 
just like the Dirac surface states of a non-magnetic TI film\cite{Pertsova2014}. 
For larger $\bf k$, the surface state bands merge with bulk bands and the 
corresponding wavefunctions are delocalized across the film\cite{SM}.
On the other hand, when the exchange field at one surface is of the order of the bulk gap, 
as in Fig.~\ref{fig1}(c), the Dirac-cone bandstructure at that surface is strongly modified; 
in particular the valence band no longer resembles a gapped Dirac cone 
even near the $\Gamma$ point. In fact, the band flattens, but remains separated from the bulk bands for most $\bf k$ values.  
The corresponding wavefunctions 
are now localized at the strongly magnetized surface for a larger region of the BZ\cite{SM}. 
For exchange fields of type (ii) (not shown in the figure),
the structure of the gapped Dirac cones is robust and, apart from the increase of the 
exchange gap with $J_{\rm t(b)}$,
remains unmodified even for $J_{\rm t(b)} > 0.3$ eV. These results demonstrate that the spatial 
distribution of $J({\bf r})$ plays a separate role from its strength in influencing 
how the Dirac surface state electronic structure is modified by surface magnetism.

\begin{figure}[ht!]
\centering
\includegraphics[width=0.98\linewidth,clip=true]{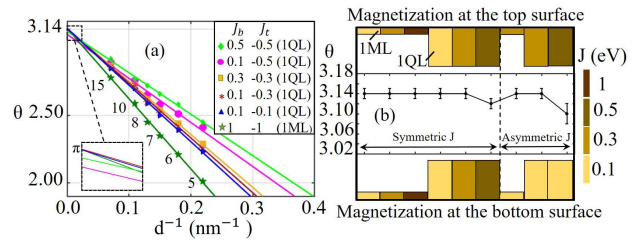}
\caption{Magneto-electric coefficient $\theta$ extracted from the calculation of the magnetization
in Bi$_2$Se$_3$ TI thin films of different thicknesses $d$
and different values of the top  and bottom surface exchange fields. (QL) and (ML) indicate that the exchange field is applied uniformly to 
the first surface QL or to the first surface monolayer respectively.
(a) $\theta$ vs $1/d$. (b) Asymptotic value ($d \to \infty$) of $\theta$.
} 
\label{fig2}
\end{figure}

We now consider the implications of these electronic structure properties for the magneto-electric response.
We compute $M_z$ numerically as a function of $E_z$, represented in the TB Hamiltonian as an on-site energy varying linearly
from the bottom to the top of the TI film\cite{SM}. 
By keeping $eE{_z}d$ smaller than the surface-state gap $\Delta$ in such a way that the $M_z$
depends linearly on $E_z$, we extract the $\theta$ parameter defined in Eq.~\ref{me_tensor}.
The results are shown in Fig.~\ref{fig2}. From Fig.~\ref{fig2}(a) we can see that, starting from 5QLs, $\theta$ versus $d$
is well described by the relation  $\theta= \theta_{d \to \infty}(1 -w/d)$\cite{wang_PRB2015}, where $w \sim 2 nm$ is
a non-universal length scale that can be extracted from this figure, and measures the localization of the side-wall current response.  
Finite-size corrections are larger than 10\% for film thicknesses below $\sim 20$nm.
Fig.~\ref{fig2}(b) demonstrates that for all choices of the exchange strength $J_{b/t}$ and position dependence, 
$\theta_{d \to \infty} = \pi$ to within numerical accuracy $\sim 1\%$. 
That is, the magneto-electric coefficient extrapolated to infinite thickness is, as expected, exactly quantized. Importantly, as shown in Fig.\ref{fig2}(a), the length scale $w$ decreases with increasing $J_{t(b)}$. Finite-size corrections are reduced when the surface magnetization is strengthened.

\begin{figure}[ht!]
\centering
\includegraphics[width=0.98\linewidth,clip=true]{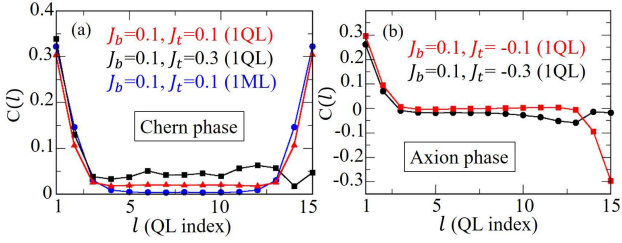}
\caption{Non-local response of Hall conductivity in 15 QL TI thin films for (a) $J_b=J_t=0.1$ and (b) $J_b=J_t=0.3$.
} 
\label{fig3}
\end{figure}

The quantization of the magneto-electric response is consistent with the properties of the non-local Hall conductivity shown in Fig.~\ref{fig3}, where 
$\sigma_{\rm H}(l,l')$ is plotted as a function of the QL indices $l$ and $l'$ for two values of the exchange constants. As anticipated, $\sigma_{\rm H}(l,l')$ is localized near the top and bottom surfaces, where TRS is broken. Furthermore, a careful numerical evaluation of 
$ \sigma_{t(b)}= \sum_{l,l'\in\ t(b)} \sigma_{H}(l,l')$ shows that the larger $J_{t(b)}$ is the more localized $ \sigma_{t(b)}$ is at the surfaces.

To shed further light on the finite-size corrections 
we define a QL-projected total Chern number ${\cal C}(l)$ 
in which the Berry curvature of each state is weighted by the projection of that 
state onto $l^{th}$ QL:
\begin{equation}
{\cal C}(l) = \frac{1}{2\pi}\int_{\rm BZ} d^2 k\, \Omega^{(l)}_{xy} ({\bf k} )\;,
\label{proj_C}
\end{equation}
where 
\begin{equation}\small 
\Omega^{(l)}_{xy} ({\bf k} ) = 
 -{2}{\rm Im}\sum_{{
\atop \scriptstyle n= {\rm oc}} \atop \scriptstyle  n'= {\rm unoc} } 
\frac{
\langle u_{n{\bf k}} |v_y({\bf k}) |u_{n'{\bf k}}\rangle \langle u_{n'{\bf k}} |{v}_x({\bf k})| u_{n{\bf k}}\rangle}
{( E_{n{\bf k}} -  E_{n'{\bf k}}) ^2} W^{(l)}_{n{\bf k}}\;.
\label{proj_Berry_curv}
\end{equation}
In Eq.~\ref{proj_Berry_curv}
$W^{(l)}_{n{\bf k}} \equiv \sum_ {s \in l} |\langle s|u_{n{\bf k}} \rangle |^2$ is
the weight of $|u_{n{\bf k}}\rangle$ when projected on
orbitals $|s\rangle$ centered at the sites $s$ of the $l^{\rm th}$ QL.
The total Berry curvature does not single out the contribution of a particular quasi-2D band. 
However, when projected on a given QL, $\Omega^{(l)}_{xy} ({\bf k} )$ and ${\cal C}(l)$ are a good measure of the contribution 
of the surface states relative to the contribution of the bulk states.
The total $\Omega$ and ${\cal C}$ obtained by summing Eqs.~\ref{proj_C}, \ref{proj_Berry_curv} over $l$,
yield $C=1$ in the Chern insulator state and $C=0$ in the axion insulator state,
regardless of the number of QLs and the value of $J$.

\begin{figure}[ht!]
\centering
\includegraphics[width=0.98\linewidth,clip=true]{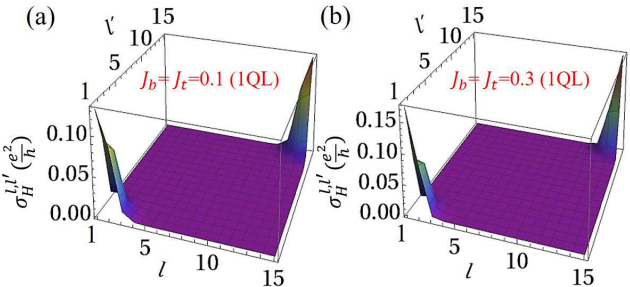}
\caption{QL-projected total Chern number for 15 QL TI thin films with different choices for the surface exchange 
fields $J_b$ and $J_t$. (a) Chern insulator configuration; (b) Axion-insulator configuration. 
The total Chern number is dominated by sums over the three QLs closest to the either surface 
only when $J_b$ and $J_t$ are both smaller than the bulk gap $\approx 0.3$ eV.
} 
\label{fig4}
\end{figure}

In Fig.~\ref{fig4} we plot ${\cal C}(l)$ versus the QL index $l$ for a 15QL film for different values of $J_{b/t}$.
Consistent with the bandstructures in Fig.~\ref{fig1}, these results show that for $J_{b/t}$ 
substantially smaller than the bulk gap, the only states with substantial Berry curvature are those that derive from the 
non-magnetic state Dirac cones, which are strongly localized in the first three QLs.\cite{SM}. 
Note that the first three QLs represent the typical localization region of the evanescent surface states in non-magnetic TI films\cite{Pertsova2014},
In this case, we can operationally define a surface QAHE conductance by 
$\tilde \sigma_{H, t(b)} \equiv e^2/h \sum_{l= t (b)}{\cal C}(l)$,
and find that it is closer to $\pm e^2/2h$; in other words the explanation for the TME in terms of surface localized half-quantized Hall conductivities 
applies literally. Only under these conditions do the definitions of a top and bottom surface Hall conductivity in terms of a 
projected Chern number ${\cal C}(l)$ and a non-local Hall conductivity $\sigma_{\rm H} (l, l')$ coincide. 
For very strong exchange potentials across a full quintuple layer our calculations show that states with large Berry curvature have substantial weight deep in the bulk of the film.\cite{SM}
This is particularly evident in the asymmetric case $J_b = 0.1$, $J_t = 0.3$ eV.  Our explicit calculation of the magnetization 
response to electric fields nevertheless shows that finite-size corrections to the magneto-electric response are actually smaller 
in this case, implying that the side-wall current response is even more concentrated in the layers of the film that have broken time-reversal symmetry, 
as evidenced by Fig.~\ref{fig3}.  For strong exchange interactions the localized side-wall response cannot be understood simply in 
terms of Hall response of states localized at the surface.

\textit{Discussion}---
We have presented a unified analysis of the QAHE, the response of side-wall current to changes in 
chemical potential, and the QTME, the response of the magnetization associated with sidewall currents to changes 
in electric potential across the width of the film.  
The QAHE and the QTME are both of fundamental importance because of the direct
relationship of these observables to Bloch state topology.
Because the electric potentials in which we are interested are 
independent of lateral position, we can introduce them by adding layer-dependent lateral electric 
fields that are non-zero only near the film side walls.  
In this way we arrive at Eq.~\ref{eq:swnl}, which
relates the magnetization and the side-wall currents to both electric potentials and chemical potentials via a Hall-conductivity 
$\sigma_H(l,l')$ that is non-local across the film.

We have argued that in the thick film limit the non-local Hall conductivity $\sigma_H(l,l') \approx 0$ when either $l$ or $l'$ are far from both the top and bottom surfaces and therefore in a region that is locally time-reversal invariant. Numerical evaluation of $\sigma_H(l,l')$  for  a  realistic TB  model of a TI thin film reported on in Fig.~\ref{fig3} support this statement, which allows us to separate the total Hall conductivity $\sigma$ into top and bottom surface contributions 
$\sigma = \sigma_t+\sigma_b$.  It follows from time-reversal symmetry that 
$\sigma_{t}(J_{t},J_{b}) = - \sigma_{t}(-J_{t},-J_{b}) $ and $\sigma_{b}(J_{t},J_{b}) = - \sigma_{b}(-J_{t},-J_{b}) $,
where $J_t$, and $J_b$ specify the exchange couplings at the 
top and bottom surfaces.  We have argued that in the thick film limit a locality condition also applies - namely that 
$\sigma_t$ depends only on $J_t$ and $\sigma_b$ depends only on $J_b$. This locality condition is certainly satisfied in the limit of weak time-reversal symmetry breaking where the exchange gap is considerably smaller than the bulk gap and the Hall conductivity is contributed by weakly gapped Dirac-cone surface states.  It is less obvious, perhaps, that the locality condition is 
satisfied in the strong $J_t,J_b$ limit where we have shown that states with large Berry curvature 
are extended across the sample - although even here non-local response would be very surprising.
The assumption of locality is supported by the fact that, when combined with quantization of the total
Hall conductivity and time-reversal symmetry properties, it naturally accounts for the expected quantization value of the 
QTME $\sigma_t=\sigma_b=e^2/2h$, which we confirm numerically,
and for the $(1-w/d)$ form of finite-size corrections, with $w$ being a 
non-locality length of the side-wall response and $d$ the film thickness.  

We have established the $(1-w/d)$ finite-size law using explicit calculations of magnetization in the presence of an electric field applied across a Bi$_2$Se$_3$ topological insulator which yield $w \sim 2 nm$.  If this is correct, the part in $10^6$ quantization accuracy routinely achieved for quantum Hall systems would require films of $\sim mm$ thickness. Stronger surface magnetism generates larger quasi-2D gaps, which in turn imply greater robustness of the quantized response against potential disorder that is inevitable and can invalidate quantization by inducing surface electron or hole puddles. It is therefore encouraging for QTME measurement efforts that finite-size corrections to the QTME are smaller for surface magnetism that is stronger. This can be achieved either in the sense of coupling more strongly to electron spins near the Fermi level or in the sense of being present over more near-surface layers of the film. 

This work was supported by the Faculty of Technology at Linnaeus University and by the
Swedish Research Council under Grant Number: 621-2014-4785.
AHM was supported by the Army Research Office under Grant Number W911NF-16-1-0472.
We acknowledges valuable interactions
with David Vanderbilt and Peter Armitage.
Computational resources were provided by the Swedish National Infrastructure for Computing (SNIC) at Lunarc partially funded by the Swedish Research Council through grant agreement no. 2018-05973.

\bibliography{QTME}

\begin{thebibliography}{38}
\bibitem [{cav()}]{caveat1}%
  \BibitemOpen
  \href@noop {} {\bibinfo  {journal} {This form assumes that the in-plane
  current and velocity projection operators commute}\ }\BibitemShut {NoStop}%
\end{thebibliography}
\end{document}